\documentclass{aa}

\usepackage[dvips]{graphicx}
\usepackage{txfonts}
\usepackage{psfig}
\usepackage[authoryear]{natbib}

\begin{document}

\title{Comparison of solar horizontal velocity fields from SDO/HMI and Hinode data}

\author{ Th.~Roudier\inst{1}, M.~Rieutord,\inst{1}, V.~Prat,\inst{1},
J.M.~Malherbe\inst{2}, N.~Renon \inst{3}, Z.~Frank\inst{4},
M.~\v{S}vanda\inst{5,6}, T.~Berger\inst{7}, R.~Burston\inst{8}, L.~Gizon\inst{8,9}}

\date{Received \today  / Submitted }

\offprints{Th. Roudier}
 
\institute{Institut de Recherche en Astrophysique et Plan\'etologie, Universit\'e de Toulouse, CNRS,
14 avenue Edouard Belin, 31400 Toulouse, France
\and LESIA, Observatoire de Paris, Section de Meudon, 92195 Meudon, France
\and CALMIP, DTSI Universit\'e Paul Sabatier, Universit\'e de Toulouse 31062 Toulouse, France
\and Lockheed Martin Advance Technology Center, Palo Alto, CA-94304, USA
\and Astronomical Institute, Faculty of Mathematics and Physics, Charles University in Prague, V Hole\v{s}ovi\v{c}k\'{a}ch 2, 
CZ-18000, Prague 8,Czech Republic
\and Astronomical Institute, Academy of Sciences of the CzechRepublic (v. v. i.), Fri\v{c}ova 298, CZ-25165, Ond\v{r}ejov, Czech
Republic
\and National Solar Observatory, Sunspot, NM 88349, USA
\and Max-Planck-Institut f\"ur Sonnensystemforschung, Max-Planck-Strasse 2, 
37191 Katlenburg-Lindau, Germany.
\and Institut f\"ur Astrophysik, Georg-August-Universit\"at G\"ottingen, 
Friedrich-Hund-Platz 1, 37077 G\"ottingen, Germany}

\authorrunning{Roudier et al.}
\titlerunning{Comparison of solar horizontal velocity maps measured using SDO/HMI and Hinode data}

\abstract{
The measurement of the Sun's surface motions with a high spatial and
temporal resolution is still a challenge.
}{
We wish to validate horizontal velocity measurements all over the visible disk of the 
Sun from Solar Dynamics Observatory/ Helioseismic and Magnetic Imager (SDO/HMI)  data.}
{
Horizontal velocity fields are measured by following the proper motions
of solar granules using a newly developed version of the Coherent Structure Tracking (CST) code.  
The comparison of the surface flows measured at high spatial resolution (Hinode, 0.1 arcsec) and low resolution 
(SDO/HMI, 0.5 arcsec) allows us to determine corrections to be applied to the horizontal velocity measured from 
HMI white light data.}
{
We derive horizontal velocity maps with spatial and temporal resolutions of 
respectively 2.5 Mm and 30 min. From the two components of the horizontal velocity $v_{\rm x}$ and $v_{\rm y}$
measured in the sky plane and the simultaneous line of sight component from SDO/HMI dopplergrams $v_D$, we derive 
the spherical velocity components ($v_r$, $v_\theta$, $v_\varphi$).  The azimuthal component $v_\varphi$ gives the 
solar differential rotation with a high precision ($\pm 0.037$~km~s$^{-1}$) from a temporal sequence of only 
three hours.
}
{
 By following the proper motions of the solar granules, we can revisit the dynamics of the solar surface 
at high spatial and temporal resolutions from hours to months and years with the SDO data.}


\maketitle
\section{Introduction}

The dynamics of the Sun's surface is one of the major elements in
understanding the time evolution of its magnetic activity. It is a
real challenge to measure the surface motions at all spatial and temporal
scales and compare them to those coming from  the simulations. Recently,
the HMI instrument aboard the SDO satellite allowed us a new step
in that direction.  By following the proper motions of the solar
granules, representative of solar plasma evolution, it is possible to
 define the flow field on the solar surface \cite[][]{Roud2012} 
 from a small spatial scale of 2.5 Mm up to nearly 85\% of the 
solar radius (Fig. 6 of that paper).  We measured the
velocities in a cartesian coordinate system, where $x$ and $y$ denote
the coordinates in the sky plane with $x$ parallel to the direction
of the solar rotation and $z$ directed towards the observer along the
line of sight. Beyond 0.85~R$_\odot$, the $v_{\rm x}$ and~ $v_{\rm y}$
components appeared to be noisy, but the $ v_{\rm x}$ component showed
a trend  indicative of the solar rotation. In order to  identify 
whether improvements could be made in the determination of the
horizontal velocities beyond that limit, we used simultaneous observations of the
Sun in white light at low (SDO/HMI) and high (Hinode) spatial resolution.

 In this paper, we describe a comparison between velocity fields projected
on the sky plane that are obtained in the south pole region with Hinode
data and SDO/HMI data using the CST code \cite[][]{RRRD07,Roud2012}. 
In the next section we discuss the corrections
done in order to get an accurate velocity close to the solar limb.
From the $v_{\rm x}$, $v_{\rm y}$ and Doppler observations, we
describe the transformation to the local surface velocities $v_r$,
$v_\theta$, $v_\varphi$. Finally, we present an application to measure
solar differential rotation with a short time sequence (3h) up to high
latitudes with low noise. Discussion and conclusions follow.

\section{Observations}

\subsection{Hinode observations}

We used data sets from the Solar Optical Telescope (SOT), onboard the
\textit{Hinode} \footnote{Launched in 2006 the Hinode spacecraft , was
designed and is now operated by JAXA (Japan Aerospace Exploration Agency)
in cooperation with NASA (National Aeronautics and Space Administration)
and ESA (European Space Agency).} mission \cite[e.g.,][]{STISO08,ITSSO04}.
The SOT has a 50~cm primary mirror with a spatial resolution of about 0.2\arcsec 
at a wavelength of 550~nm. For our study, we used blue
continuum observations at 450.45~nm. from the \textit{Hinode}/SOT BFI
(Broadband Filter Imager).  The observations were recorded continuously
on December 10, 2011, from 16:11:33 to 19:07:05 UT.  To get the limb
as reference, the south pole was observed at the position reported
in the Flexible Image Transport System (FITS) header of Hinode, i.e, 
$X_{\rm cent}=-10.08\arcsec$ and $Y_{\rm cent}=-969.84\arcsec$.  
The time step was 45 sec and  the field
of view was $111.6~\arcsec\times111.6~\arcsec$ with a pixel size of
0 \farcs1089. ~After alignment, the useful field of view reduced to
$104\arcsec\times82\arcsec$.  To remove the effects of the oscillations,
we applied a subsonic Fourier filter.  This filter was defined by a
cone in $k$-$\omega$  space, where $k$ and $\omega$ are spatial and
temporal frequencies. All Fourier components such that $\omega/k\geq
V_{\rm cut-off}=7\,\mathrm{km~s}^{-1}$ were removed so as to keep only 
convective  motions \cite[][]{TTTFS89}.

\subsection{SDO/HMI observations}

The HMI \cite[][]{Scherrer2012,Schou2012} onboard the Solar SDO provides 
uninterrupted observations over the entire disk. This gives a unique 
opportunity for mapping surface flows on various scales (spatial and 
temporal). Using the SDO/HMI white light data on December 10, 2011, from 
16:11:15 to 19:06:45 UT, we derived horizontal velocity fields from image 
granulation tracking using the newly developed version of the CST code 
\cite[][]{Roud2012}. The time step was 45 seconds with a pixel size of 
0\farcs5. The solar differential rotation discussed in Sect. 6 was 
determined from SDO/HMI white light and Doppler data taken on August 30, 
2010 from 8:00:45 to 11:09:45 UT.

\begin{figure}[t]
\centerline{\psfig{figure=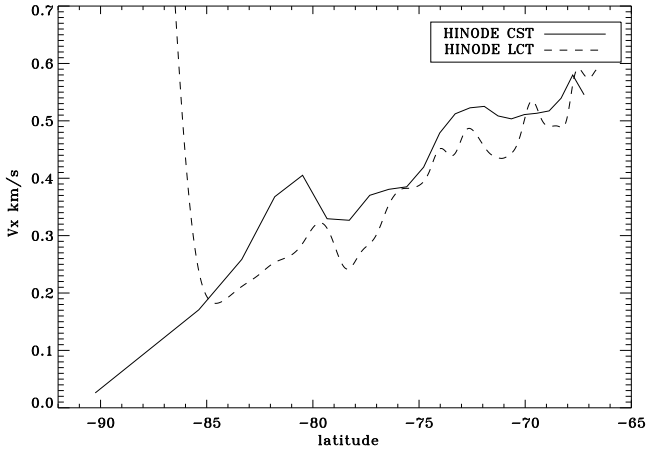,width=9 cm}}
\caption[]{$v_{\rm x}$ computed from LCT and CST on Hinode observations
on December 10, 2011.}
\label{Vx3h}   
\end{figure}

\begin{figure}[t]
\centerline{\psfig{figure=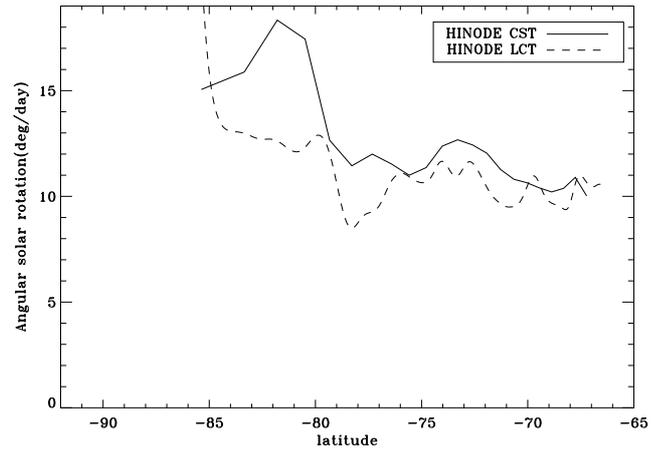,width=9 cm}}
\caption[]{ Angular rotation (sidereal) computed from LCT and CST on
Hinode observation on December 10, 2011.}
\label{Omega3h}
\end{figure}

\section{Hinode south pole flow field}
 
Measuring the solar rotation near the poles is a difficult task because there 
are few solar structures at high latitudes suitable for tracking. However, 
the Hinode images allow us to follow solar granules with good contrast up to 
the solar limb. First, we compare results of two methods to measure the solar 
motion close to the south pole using  Hinode observations: local correlation
tracking (LCT) \cite[][]{NS88} and CST \cite[][]{RRRD07,Roud2012}. 
Both methods track the horizontal motions of granules in
the field of view. More precisely, the LCT method of obtaining the horizontal 
velocity uses a spatial window that simultaneously accounts for the solar granules 
and integranular structures and may cover several granules. In contrast, the CST 
method measures the velocities by following the  trajectory of each granule, i.e., 
solar plasma,  during the life of the coherent object, which is defined by its 
appearance and disappearance if the granule does not split or merge. 
The data were aligned and the P angle evolution corrected by 0.018\degr/hour 
in order to get a perfect co-alignment with SDO data used in the following. 
On December 10, 2011, the $B_0$ angle was ~$-0.25\degr$, resulting in insignificant
projection effects with a negligible evolution during the observation period. 
Figure~\ref{Vx3h} shows good agreement of the $v_{\rm x}$ component from the LCT and 
CST up to  latitude  75\degr. Close to the limb, the effect of the spatial window 
used in the LCT prevents correct velocity determination. Figure~\ref{Omega3h} gives 
a solar sidereal rotation rate close to the pole of  around 11\degr/day, which 
corresponds to a polar period of 32.73 days. This is in agreement with previous 
determinations \cite[][]{Beck2000}.

\section{Comparison of the velocities from SDO and Hinode}

Our main goal is to compare the surface flows measured with a high
spatial resolution (Hinode, 0.1\arcsec) and a low resolution (SDO,
0.5\arcsec). We extracted the same field of view from the SDO data set as
the Hinode one and performed a very precise co-alignment of both
sequences. To get the best co-alignment, the south polar region was
observed and the solar limb  was taken as an absolute reference. An
additional check was performed after the alignment process by locating
the  brightest features (facular points) at the beginning and end of
both sequences (Hinode and SDO). A very good match of the structures
indicated very good alignment of both sequences during the 3 hours. We
then applied the CST to both sequences to get the horizontal velocities
($v_{\rm x}$ and $v_{\rm y}$) over the entire field of view. Figures~\ref{Vx2h}
and ~\ref{Omega2h} show a good agreement up to 78\degr of southern
latitude. This indicates that SDO observations can be used to determine
the solar rotation in the central meridian region up to significantly
high latitudes.  However, as shown in Fig.~\ref{vy2h}, the meridional
component measured on SDO data exhibits an offset of 0.4 ~km~s$^{-1}$
relative to the Hinode component, which was used as the reference (because the velocity
is close to zero at the limb). Figure~\ref{vy2h} shows a good correspondence
between Hinode and SDO velocities when the 0.4~km~s$^{-1}$ offset is
removed (curve marked as SDO$-$0.4). We tried to elucidate the origin
of that offset by determining all possible errors as listed in Table II
of \cite{Strous2000b}. The estimation on the SDO measurement gives
a total error on the $v_{\rm y}$ component of around 0.034~ km~$s^{-1}$,
which is ten times smaller than the measured offset.

\begin{figure*}
\centerline{\resizebox{9cm}{9cm}{\includegraphics{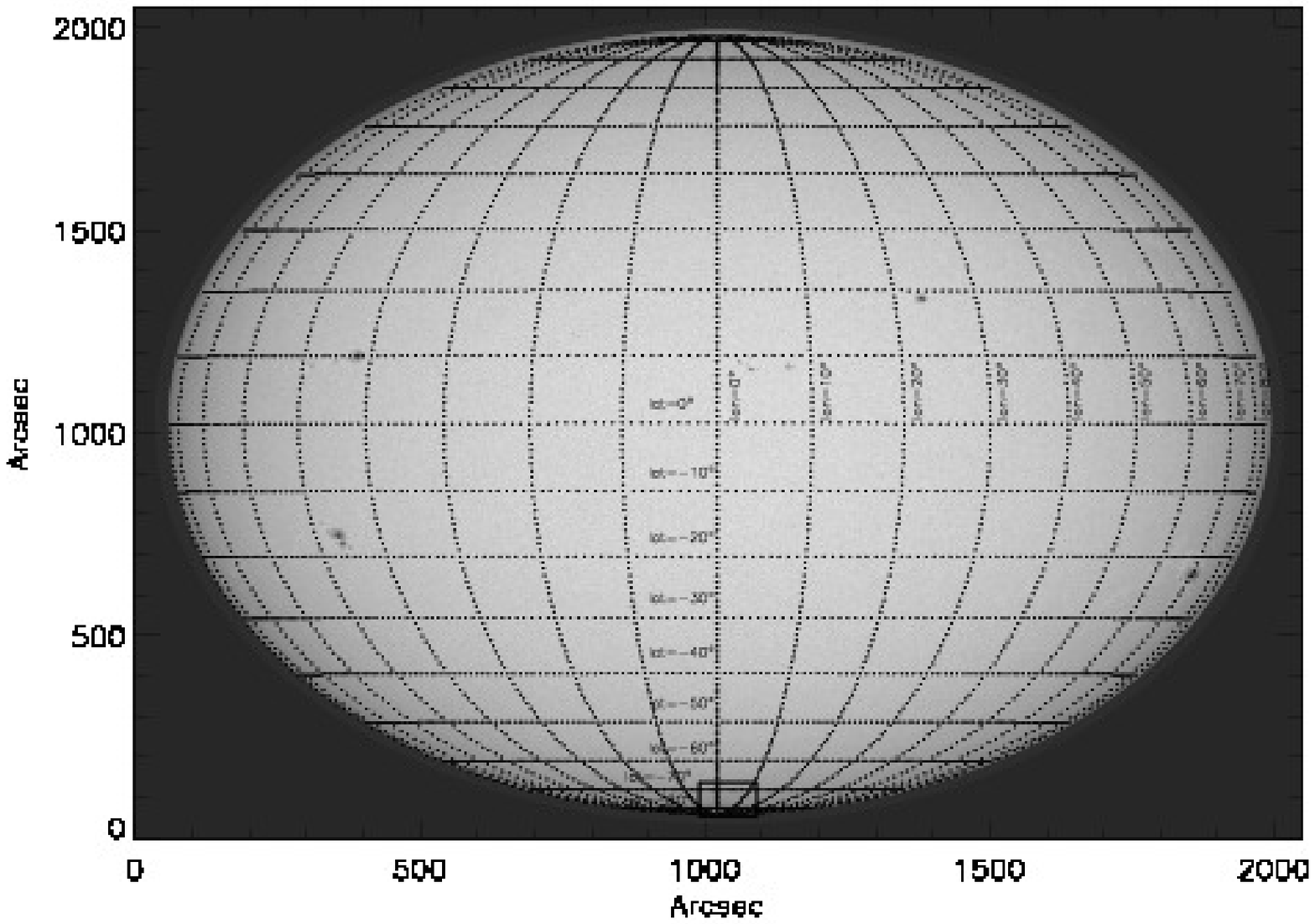}}
\resizebox{9cm}{9cm}{\includegraphics{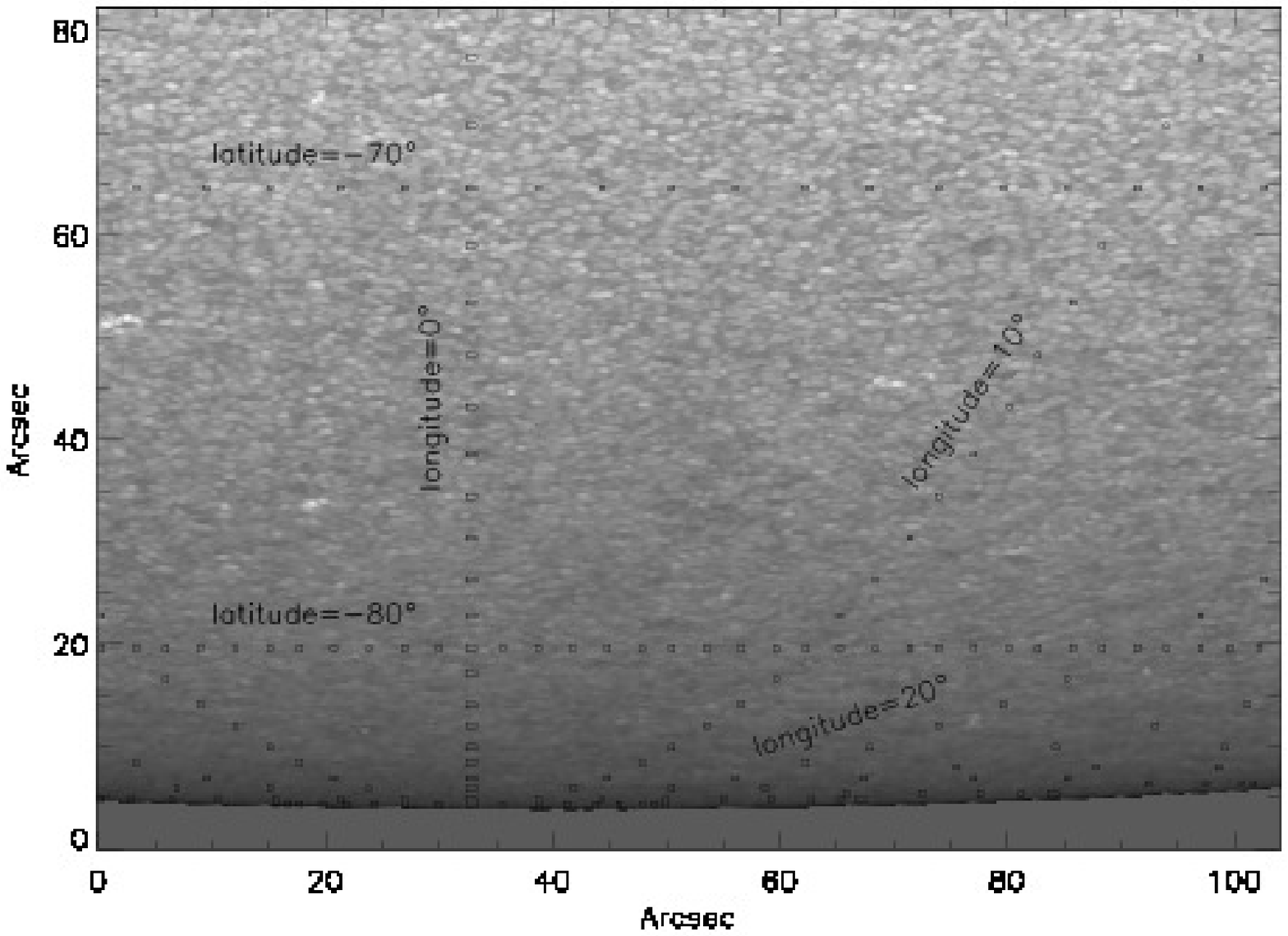}}}
\caption[]{SDO (left) and  Hinode (right) observations on December 10, 2011, at 
16:11:15 and 16:11:33 respectively. The location of the Hinode field is shown on 
the SDO image}
\label{context7}
\end{figure*}

\begin{figure}
\centerline{\psfig{figure=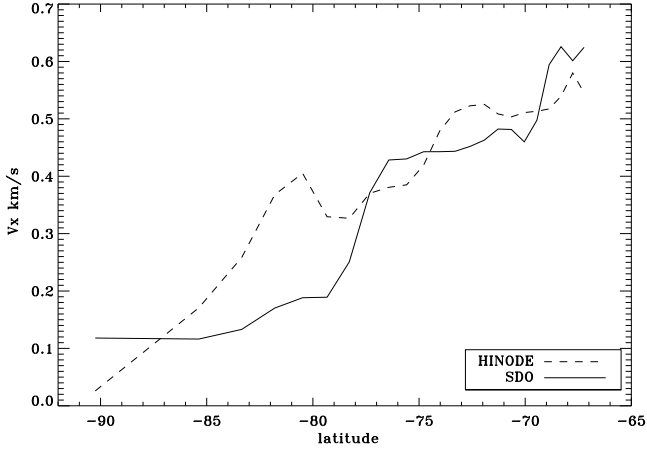,width=9 cm}}
\caption[]{$v_{\rm x}$ component from Hinode and SDO  Hinode
(0.1\arcsec) and SDO (0.5\arcsec), same field and duration.}
\label{Vx2h}
\end{figure}

\begin{figure}
\centerline{\psfig{figure=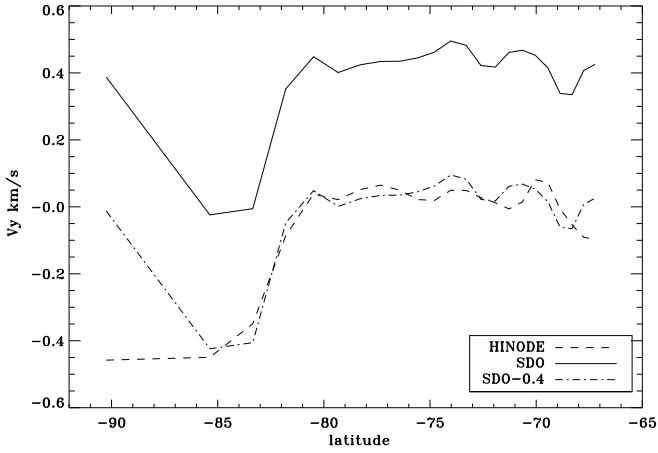,width=9 cm}}
\caption[]{$v_{\rm y}$ component from Hinode (0.1\arcsec) and SDO
(0.5\arcsec) same field and duration.}
\label{vy2h}
\end{figure}

\begin{figure}[h]
{\psfig{figure=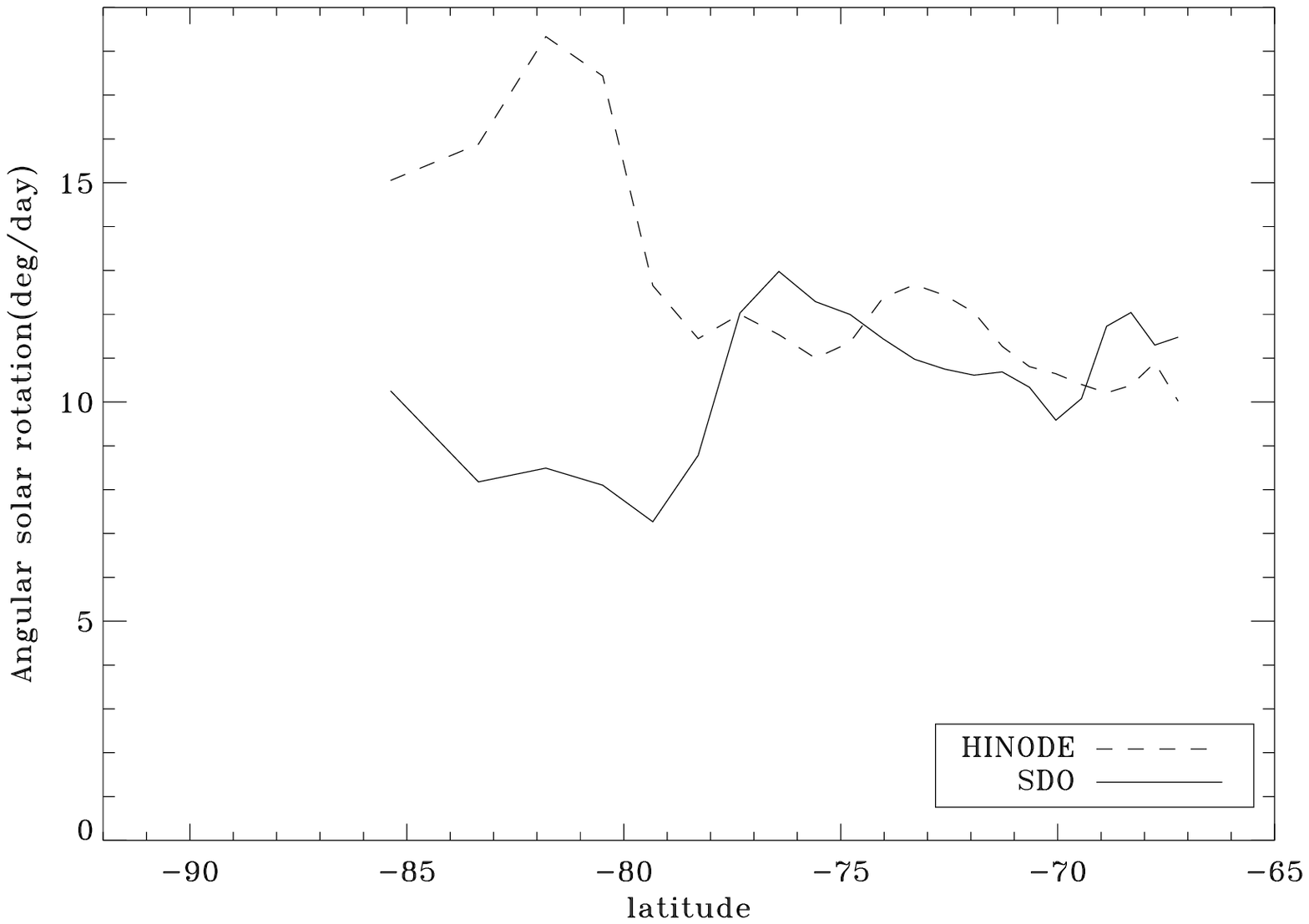,width=0.77\linewidth}}
\caption[]{Angular solar rotation from Hinode and SDO  Hinode
(0.1\arcsec) and SDO (0.5\arcsec), same field and duration.}
\label{Omega2h}
\resizebox{0.8\hsize}{!}{\includegraphics{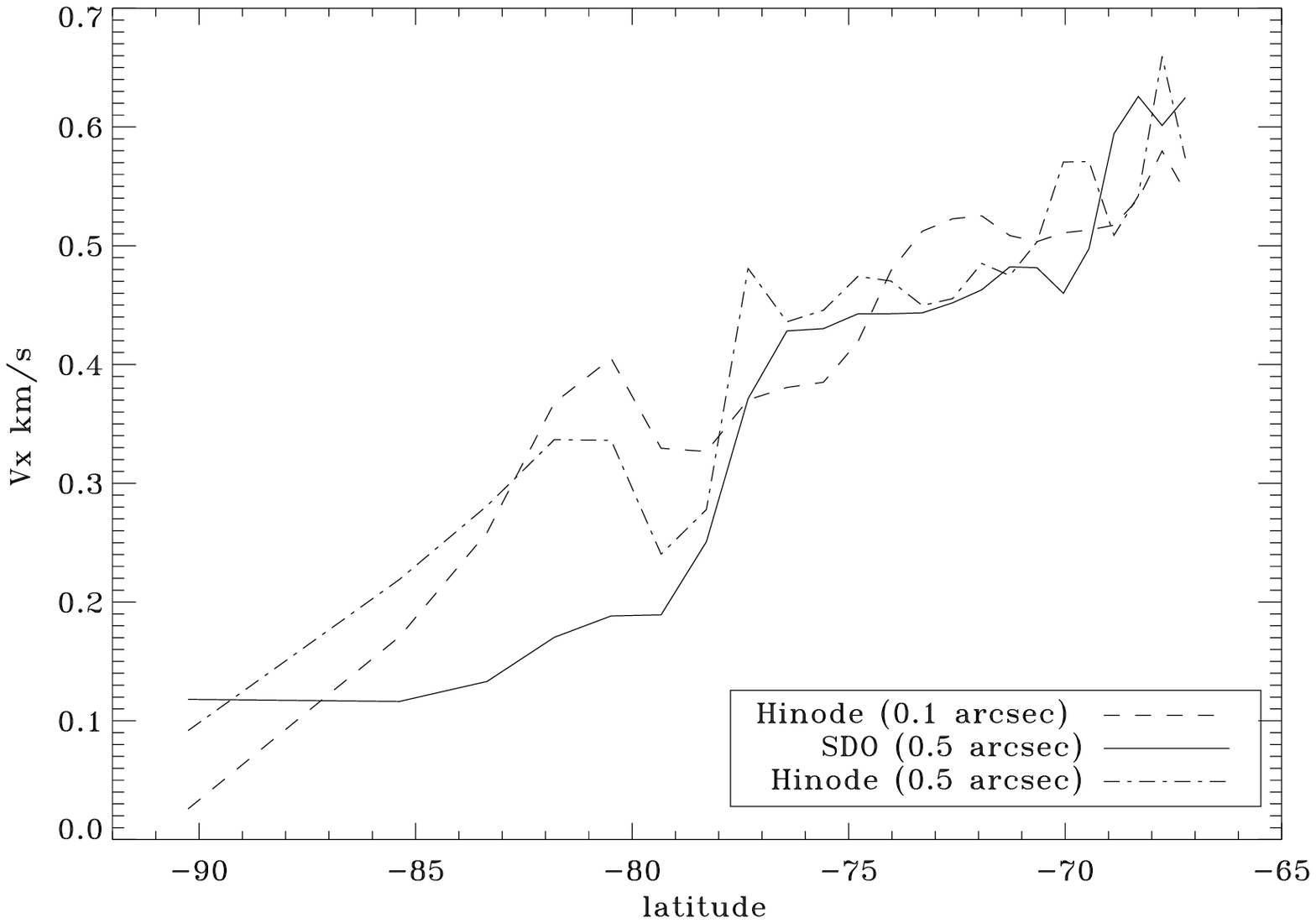}}
\resizebox{0.8\hsize}{!}{\includegraphics{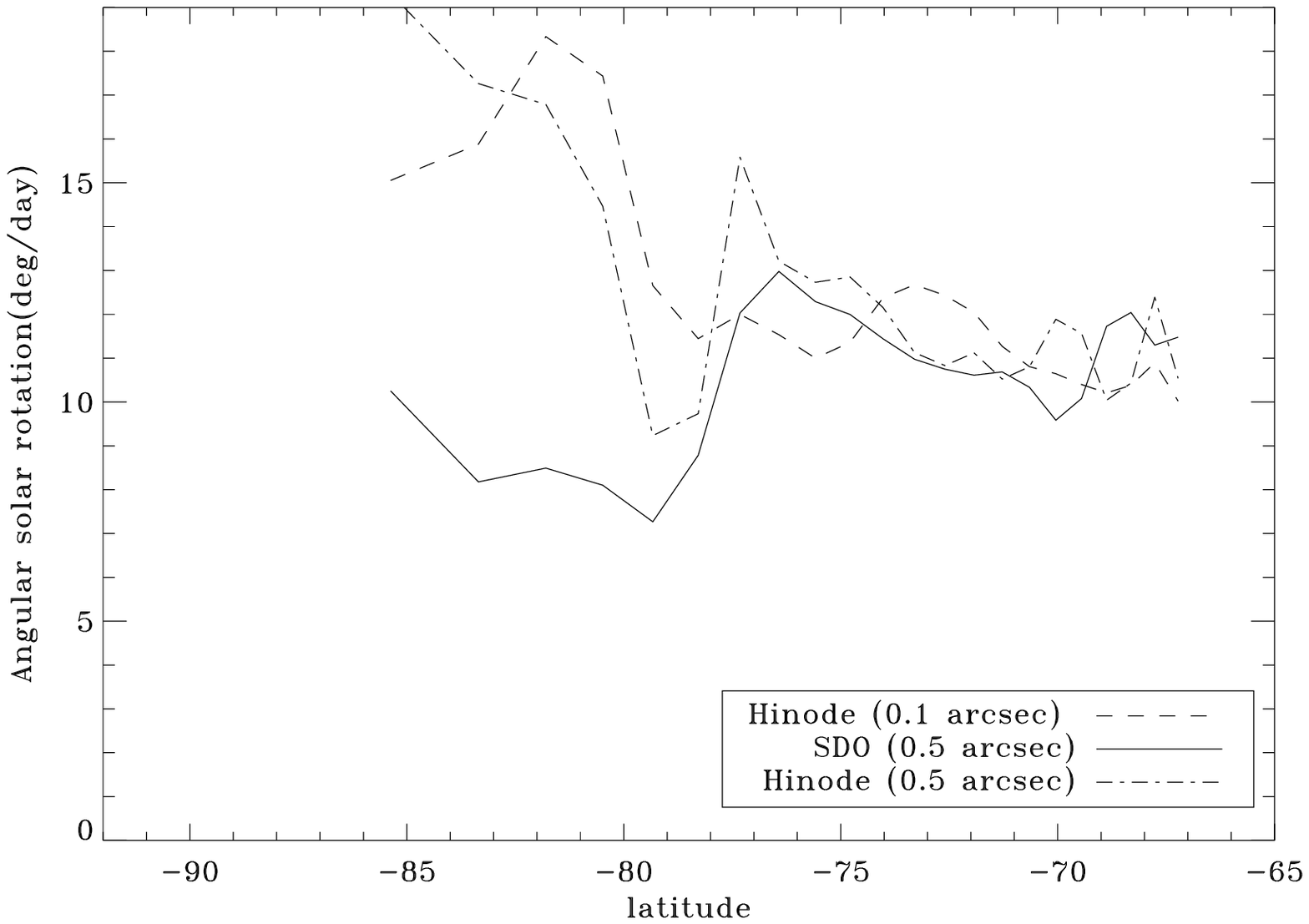}}\\
\resizebox{0.8\hsize}{!}{\includegraphics{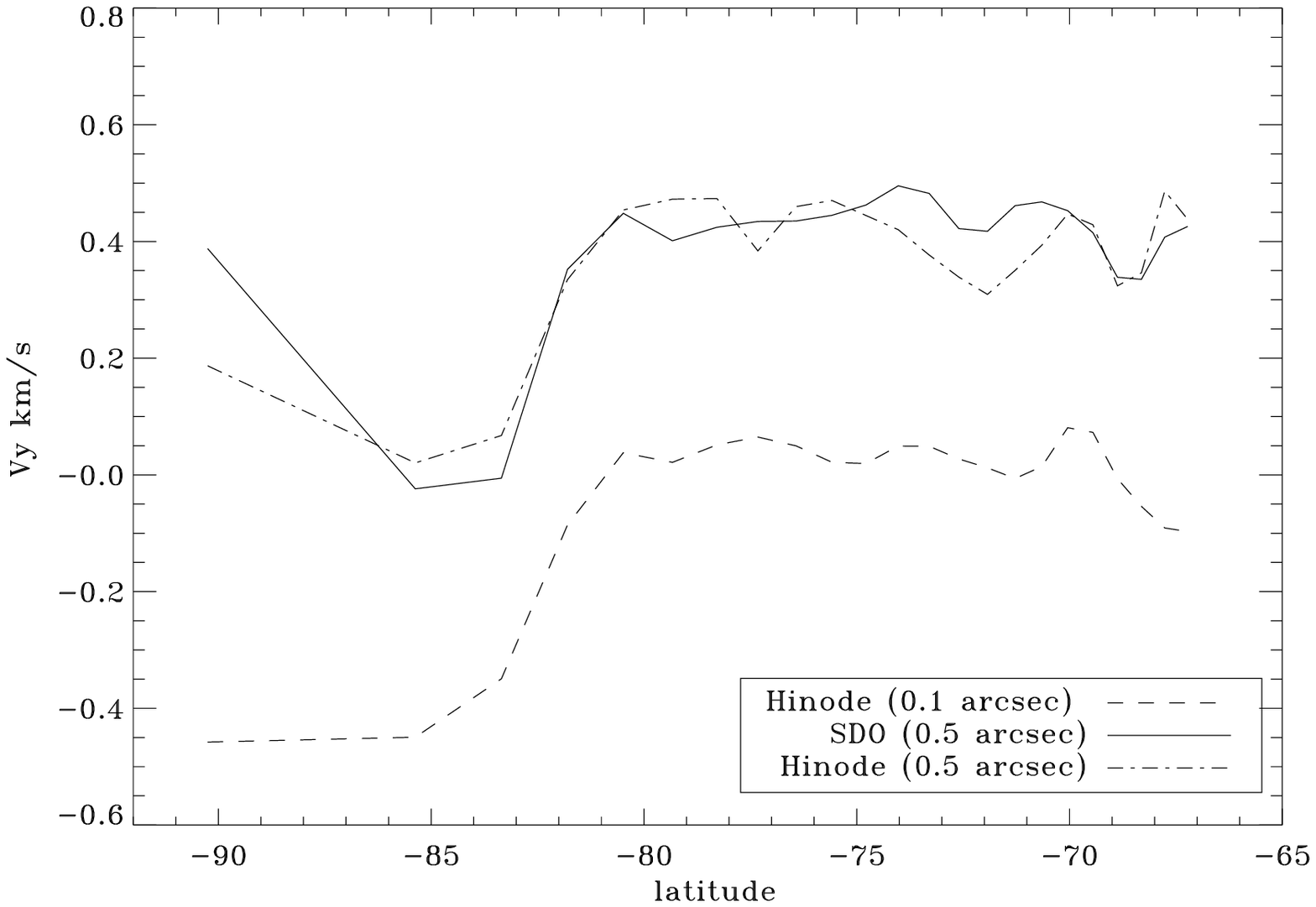}}
\caption[]{The comparison of  $v_{\rm x}$, angular solar rotation and  $v_{\rm y}$ and from Hinode (0.1\arcsec), 
Hinode (0.5\arcsec)  and SDO (0.5\arcsec) }
\label{compar}
\end{figure}

\begin{figure}
\centerline{\psfig{figure=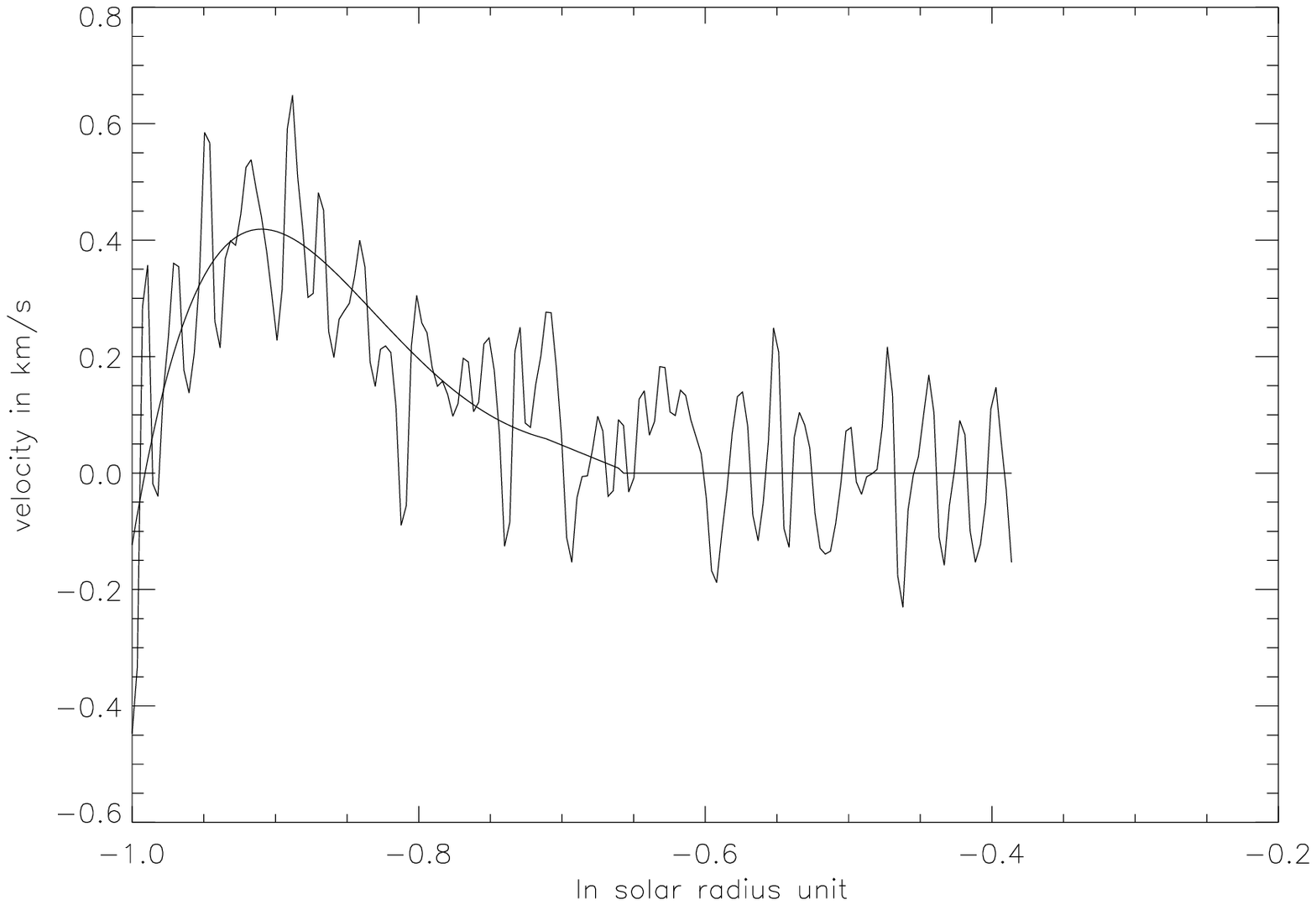,width=9 cm}}
\caption[]{ Determination of the radial offset close to the south solar 
limb and the adjusted function}
\label{correction2}
\end{figure}

\begin{figure}
\centerline{\psfig{figure=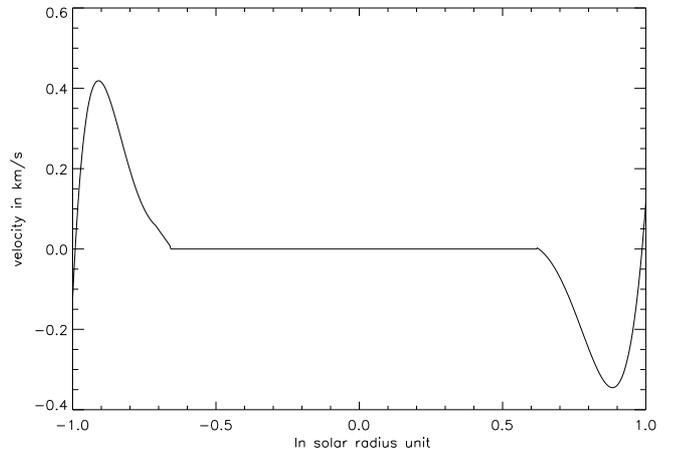,width=9 cm}}
\caption[]{Entire profile of the correction of the radial velocity to be 
applied on the velocities $v_{\rm x}$ and $v_{\rm y}$ all over the Sun }
\label{correction}
\end{figure}

As described above, one of the major differences between the two
sequences is the pixel size, which is 0.1\arcsec\ and 0.5\arcsec\ for
Hinode and SDO, respectively. In order to analyze the sequences under
the same conditions, we degraded the pixel size of the Hinode observation
to the SDO one: 0.5\arcsec.  We then applied CST to the Hinode degraded
sequence to get the horizontal velocities, which were also $k$-$\omega$
filtered. Figure~\ref{compar} still shows good agreement for the $v_{\rm
x}$ component and, in the same way, for the siderial angular rotation, at
the latitudes up to 78\degr.  The offset of 0.4~km~s$^{-1}$ observed
in $v_{\rm y}$ component of the Hinode data at 0.5\arcsec\ is clearly
visible in Fig.~\ref{compar}. We conclude that the offset is caused by
the combination of the lower spatial resolution and decreasing contrast
to the limb. One has to bear in mind that we only observe the south pole
region in detail, where  $v_{\rm y}$ is practically identical to the
radial component of the flow. Thus, our observation can be generalized
so that the radial component is always affected by the offset and we
can correct for it. Indeed, different processes play a role in generating
that radial component: at very high heliocentric angles, one pixel covers
several granules. The granule part close to the limb is of lower
contrast: it tends to be lost in the segmented images. This introduces
an artificial radial motion of some granules that we see as the
offset in the $v_{\rm y}$ component on the central meridian when comparing
high-resolution (Hinode) and low-resolution (SDO) measurements. This
radial effect was previously observed  in our measurements, but its origin
was not identified. Since we now have the origin of that error,
we can measure it and correct for it all over the Sun. One way to get
the radial offset is to average the radial velocity component over a
circle centered on the solar disk.  Due to the $B_0$ angle, the radial
correction is not necessarily symmetrical in the northern and southern hemispheres
of the Sun. This is why we treat the average process in the northern
and southern regions of the Sun separately. Figure~\ref{correction2}
shows, for example, the plot of the measured average radial component in
the southern part and the  overplotted fit obtained by using a polynomial
function of the fifth degree.  Figure~\ref{correction} shows the entire
profile of the correction of the radial velocity to be applied to the
velocities $v_{\rm x}$ and $v_{\rm y}$ all over the Sun (north and south),
allowing a good representation of the velocity field around 90\% of the
solar radius. In the following, this correction is applied systematically
to all velocity fields.

.

\begin{figure*}
\resizebox{7.0cm}{!}{\includegraphics{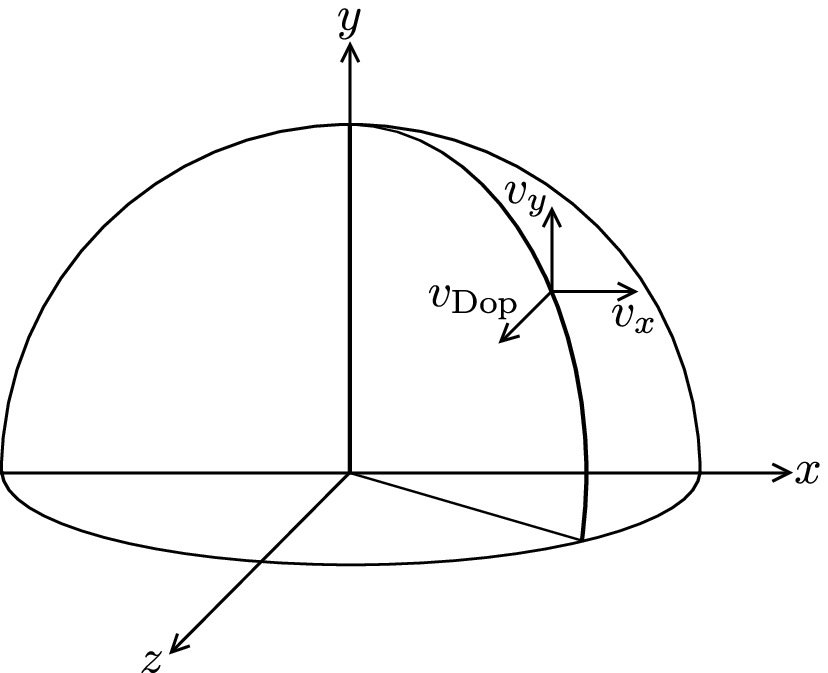}}\hspace*{2cm}
\resizebox{7.0cm}{!}{\includegraphics{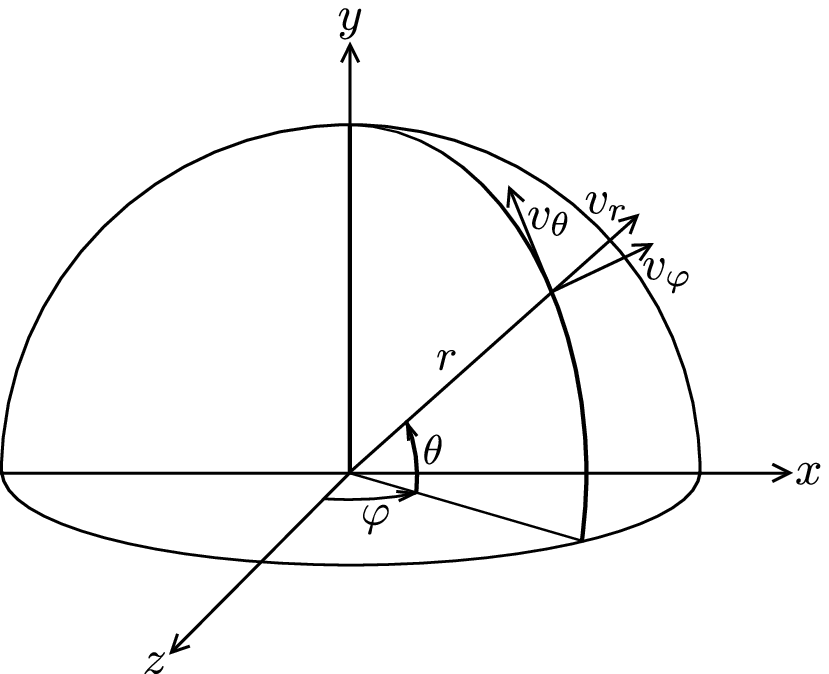}}\\
\caption[]{Coordinate systems used throughout this paper : velocity
components in the sky plane $v_{\rm x}$ and $v_{\rm y}$ and the
line of sight velocity  $v_{\rm Dop}$ (left);  Velocity components on the
solar surface $v_r, v_\varphi, v_\theta$ (right).}
\label{coordonnees}
\end{figure*}

\begin{figure*}
\resizebox{11.2cm}{!}{\includegraphics{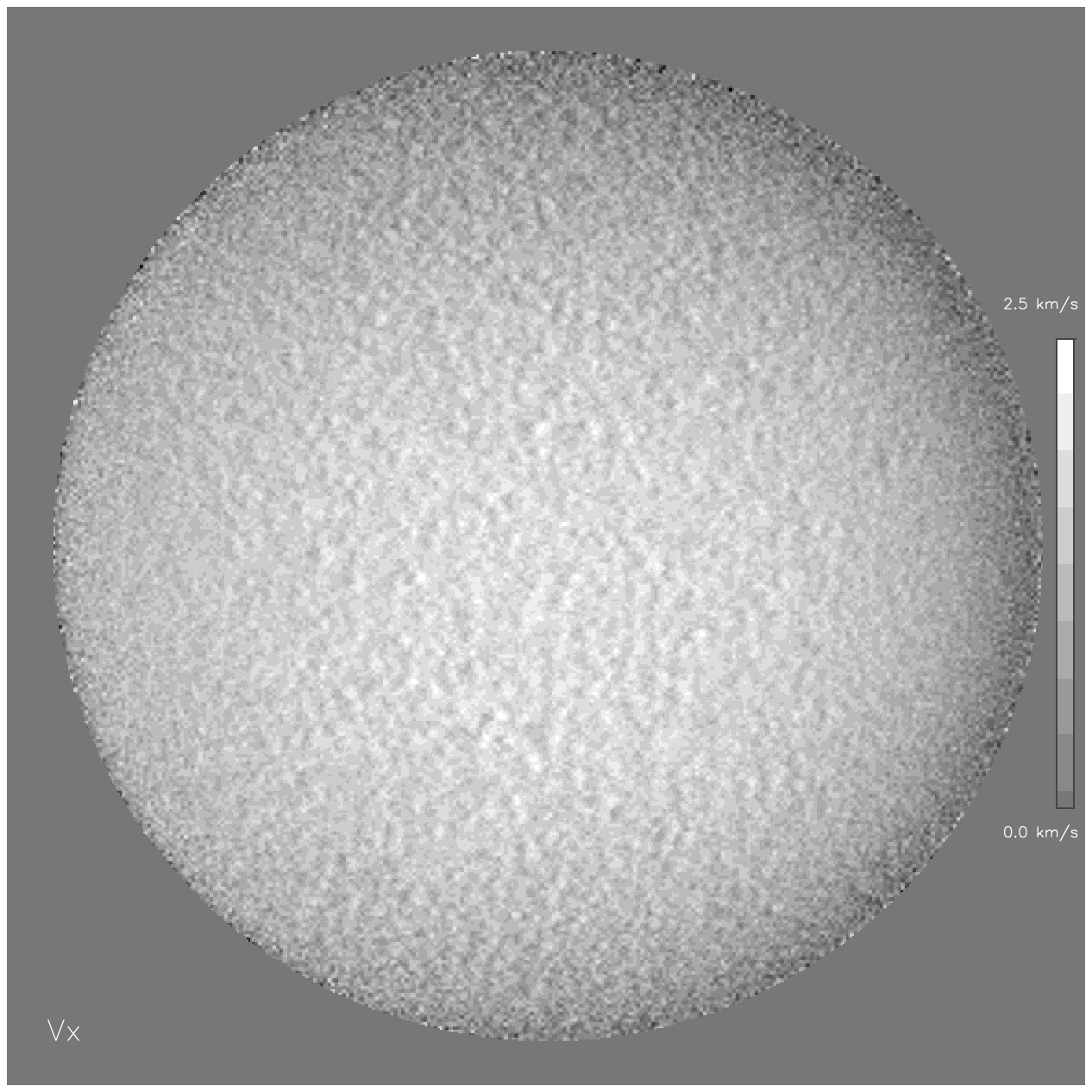}}
\resizebox{11.2cm}{!}{\includegraphics{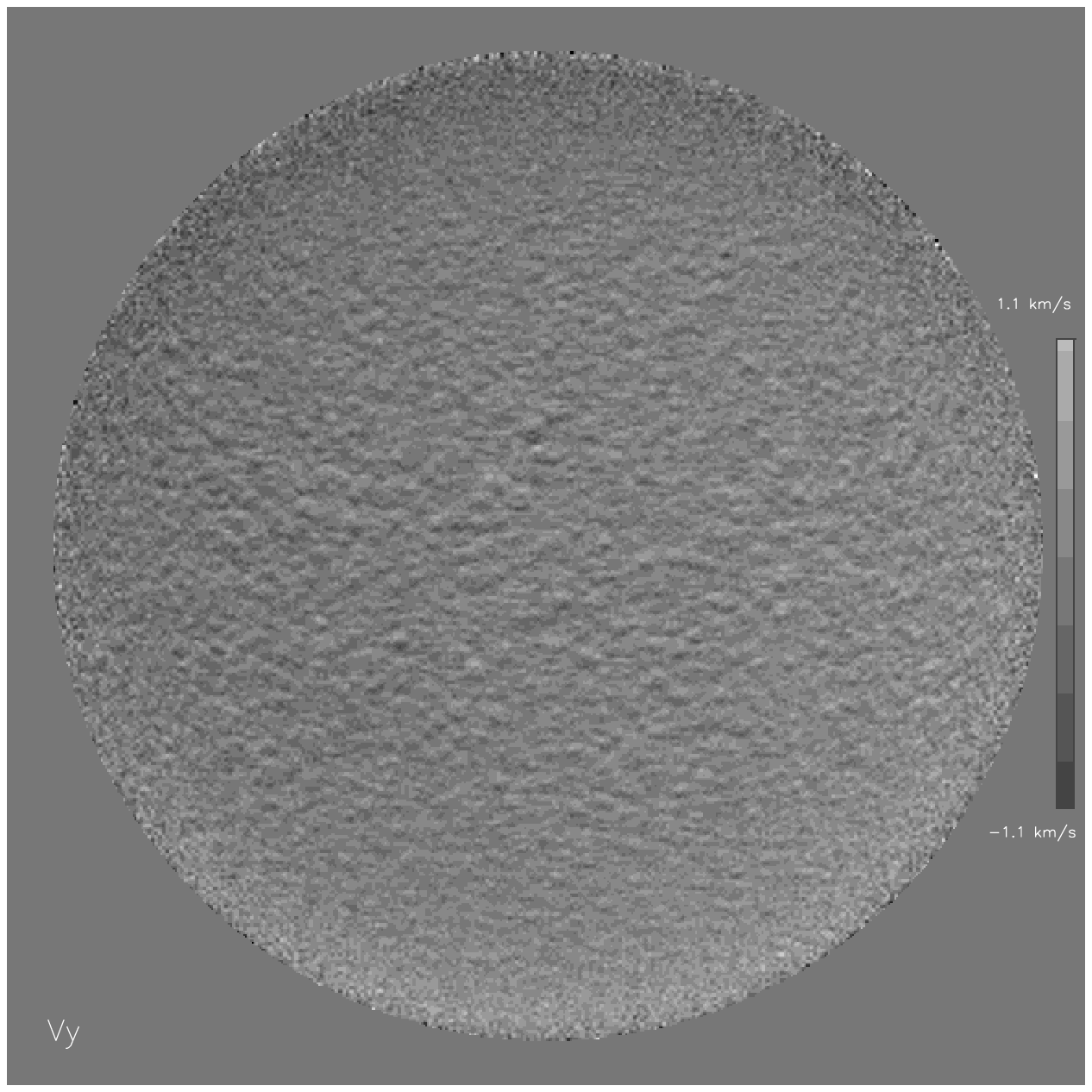}}\\
\resizebox{11.2cm}{!}{\includegraphics{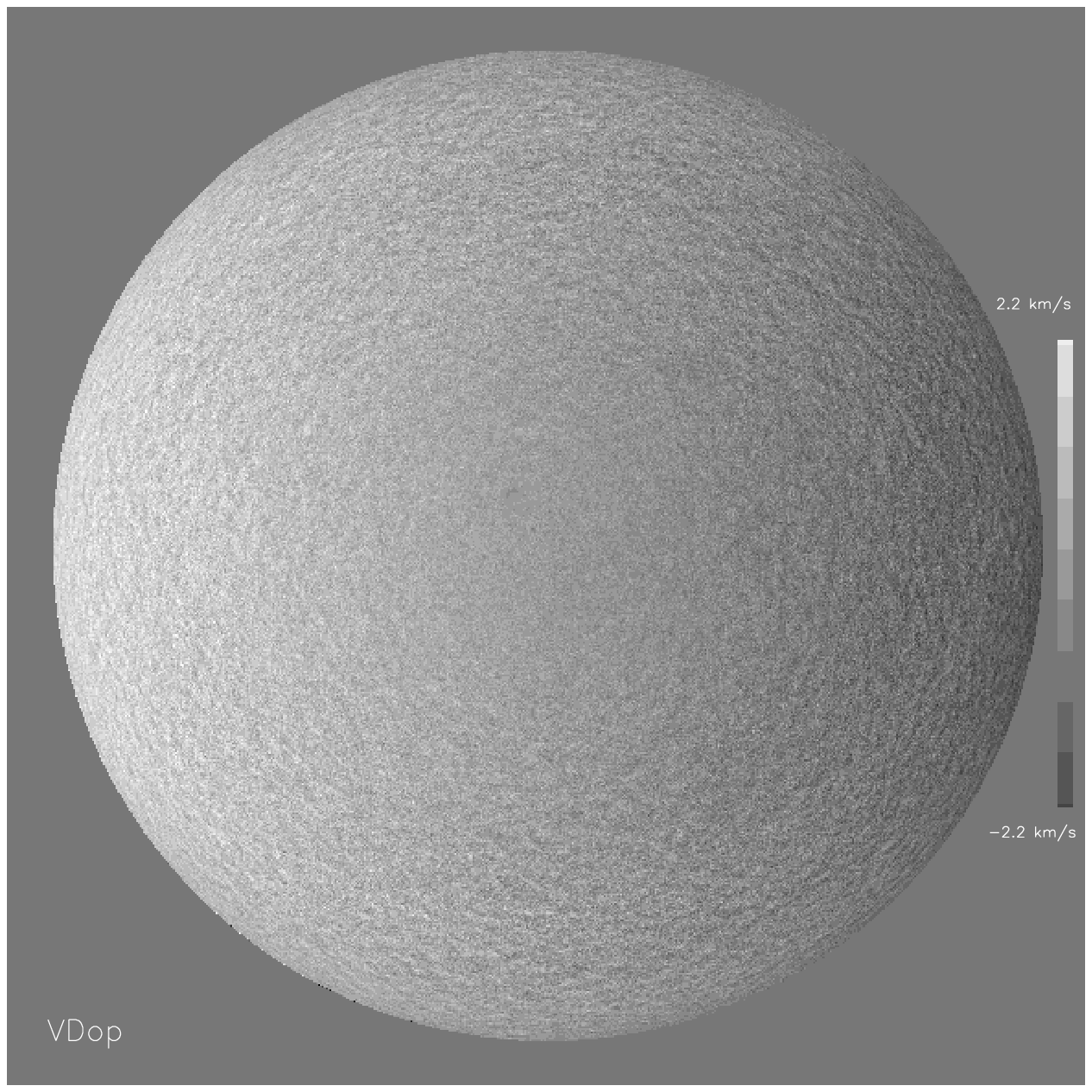}}
\resizebox{11.2cm}{!}{\includegraphics{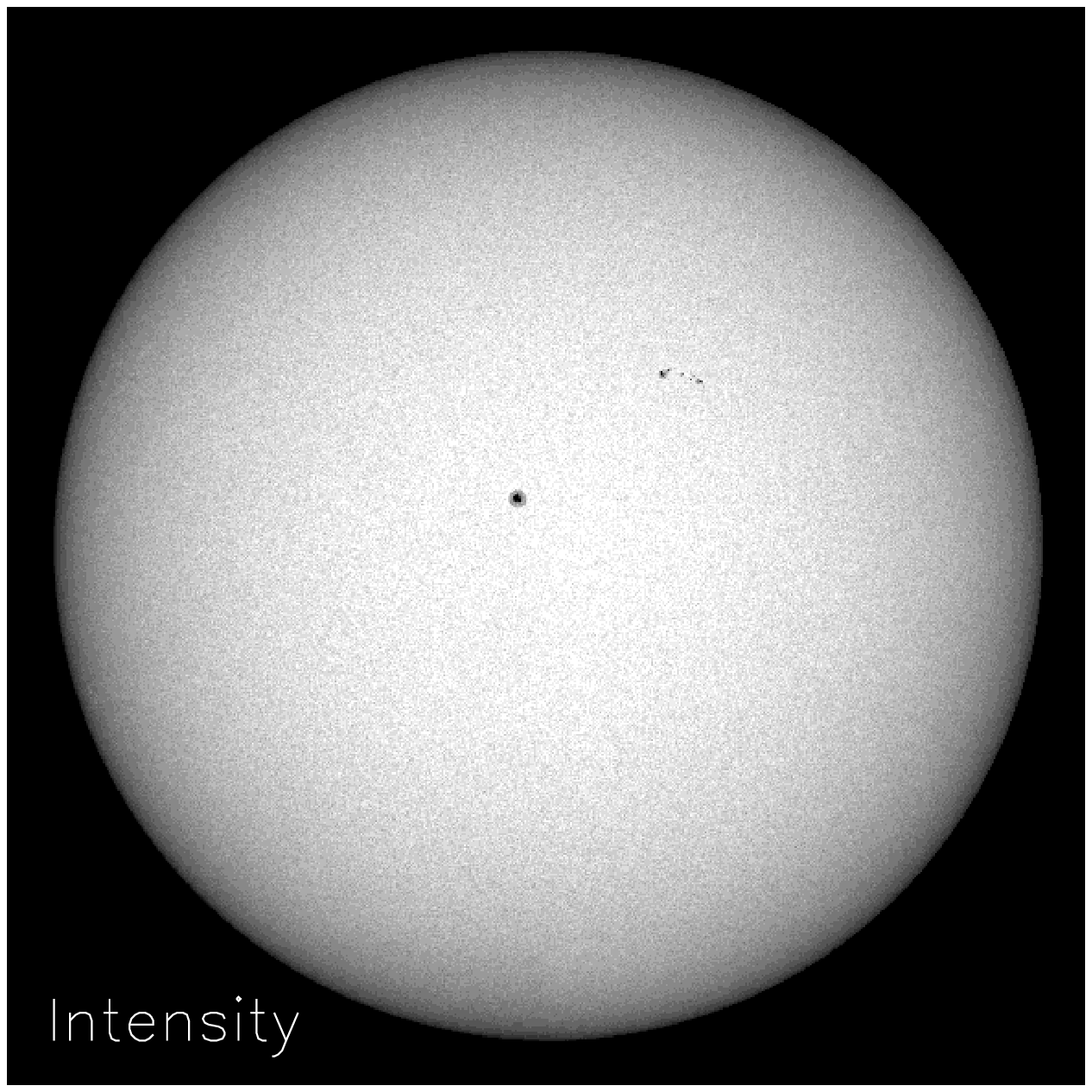}}\\
\caption[]{$ v_{\rm x}$, $v_{\rm y}$, $v_{\rm Dop}$ velocity maps for
the three-hour sequence on August 30,2010, and the Intensity at the
beginning of the sequence.}
\label{original}
\end{figure*}

\section{Determination of the spherical components of the velocity}

From the velocity components $v_{\rm x}$ and $v_{\rm
y}$ measured in the sky plane and the simultaneous line of sight velocity
$v_D$ measured from the SDO/HMI dopplergrams (e.g., Fig.~\ref{original}), we can
derive the spherical velocity components $v_r$,  $v_\theta$, $v_\varphi$,
projected onto spherical coordinates $r$, $\theta$, $\varphi$ using

\begin{eqnarray}    
v_r(\theta,\varphi) & =& \cos\theta * \sin\varphi * v_{\rm x} \nonumber\\
&+& (\sin \theta * \cos B_0 - \cos \theta * \cos \varphi * \sin B_0)* v_{\rm y}\nonumber\\
&+& (\cos \theta * \cos \varphi * \cos B_0 + \sin \theta * \sin B_0) *  v_{\rm Dop}\nonumber\\
v_\theta(\theta,\varphi) &=& -\sin\theta * \sin\varphi * v_{\rm x}  \nonumber\\
&+& (\sin \theta * \cos \varphi * \sin B_0 + \cos \theta * \cos B_0 ) * v_{\rm y}  \nonumber\\
&+& (\cos \theta * \sin B_0 - \sin \theta * \cos \varphi * \cos B_0 ) * v_{\rm Dop} \nonumber\\
v_\varphi(\theta,\varphi) &=& \cos \varphi * v_{\rm x} \nonumber\\
&+& \sin \varphi * \sin B_0 * v_{\rm y} \nonumber\\
&-& \sin \varphi * \cos B_0 * v_{\rm Dop}\nonumber
\end{eqnarray}

The system of coordinates is also shown in Fig.~\ref{coordonnees}. We note
that $\varphi$ is along the longitude and $\theta$ is along the
latitude. The co-alignment of all components has been checked carefully
by using the information from FITS headers of the SDO
data. Before using the dopplergrams, some data reduction had to be applied.
First, the dopplergrams were averaged over the same time interval as
the $v_{\rm x}$ and $v_{\rm y}$ sequence, here three hours. The limbshift had to be
be corrected \cite[][]{Ulrich2010} and, in this case, we used the limbshift function
determined by the SDO team over the month of August 2010 (P. Scherrer,
private communication).

By defining $z=1- \cos \rho$, where  $\rho$ is heliocentric angle, the
limb-shift correction (in km~s$^{-1}$) is  given by limb-shift$(z)=
-0.664 z + 0.775 z^2 + 0.284 z^3$.  The mean velocity of the central
region of the dopplergrams was taken as zero origin. The dopplergrams were
resampled to the same size as the $v_{\rm x}$ and $v_{\rm y}$ components.

To convert the $v_{\rm x}$ and $v_{\rm y}$ velocities in the sidereal system
$v_x^{\rm sid}$ and $v_y^{\rm sid}$, the Earth's orbital displacement
was taken into account.  This correction depends on the
P angle as

\begin{eqnarray}    
v_x^{\rm sid} &=&  v_{\rm x} +  v_{\rm e}\cos P \nonumber\\
v_y^{\rm sid} &=& v_{\rm y} + v_{\rm e}\sin P \nonumber
\end{eqnarray}
On August 30, 2010, it was 0.13~ km~s$^{-1}$ on
$v_{\rm x}$ and  0.04~ km~s$^{-1}$ on $v_{\rm y}$, where $v_{\rm e}$
is the Earth velocity component projected on the Sun's surface.

\begin{figure}
\begin{center}
\resizebox{\hsize}{!}{\includegraphics{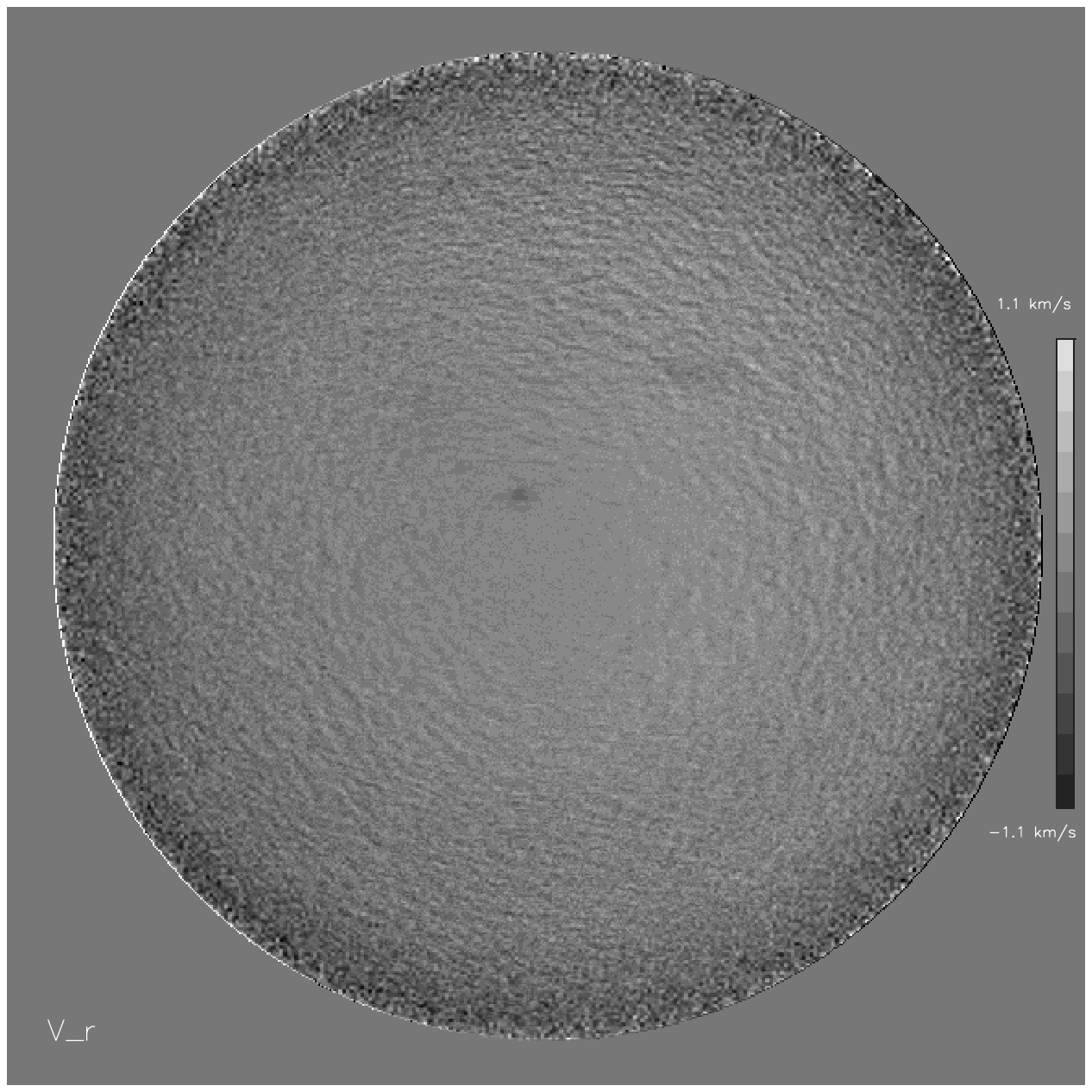}}
\resizebox{\hsize}{!}{\includegraphics{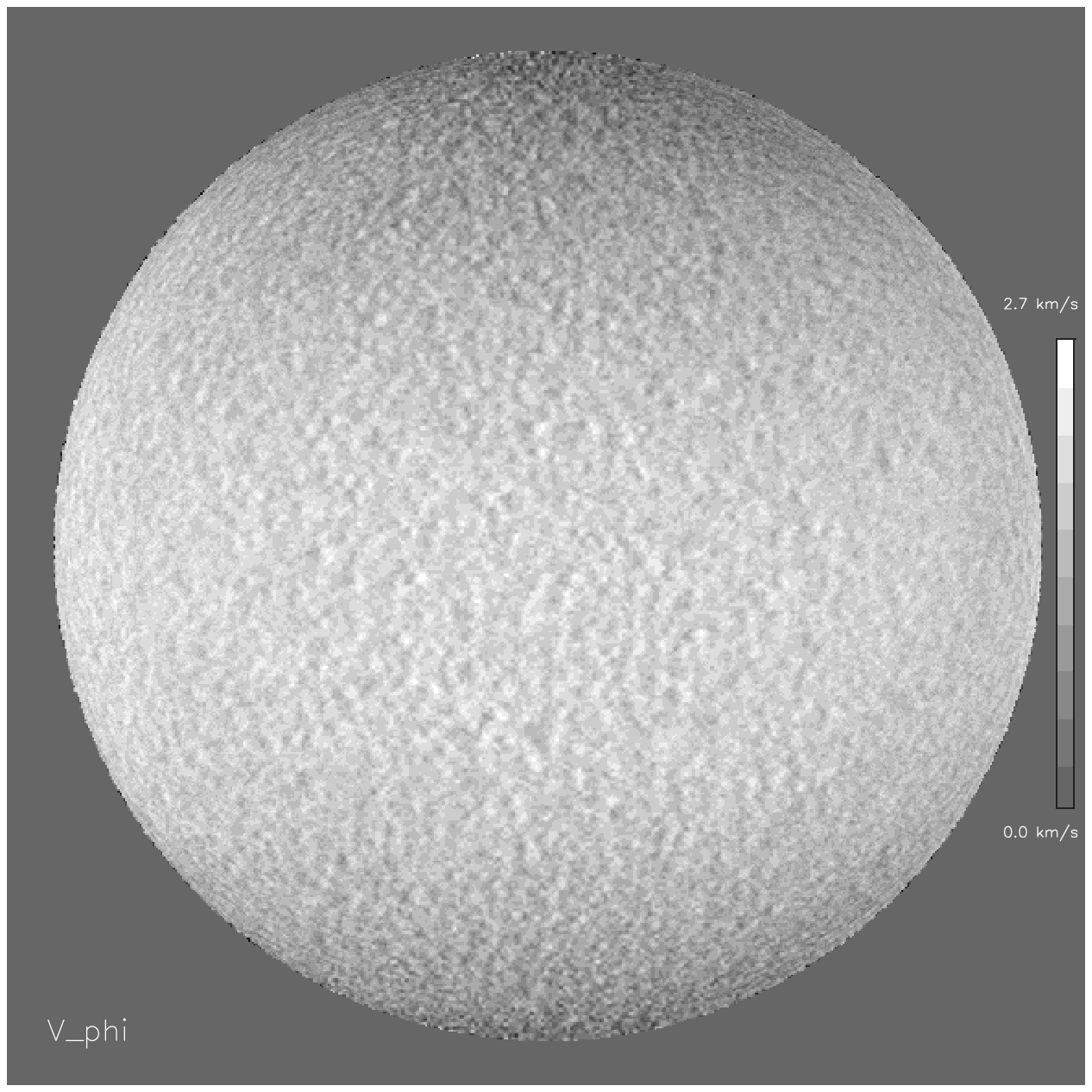}}\\
\resizebox{\hsize}{!}{\includegraphics{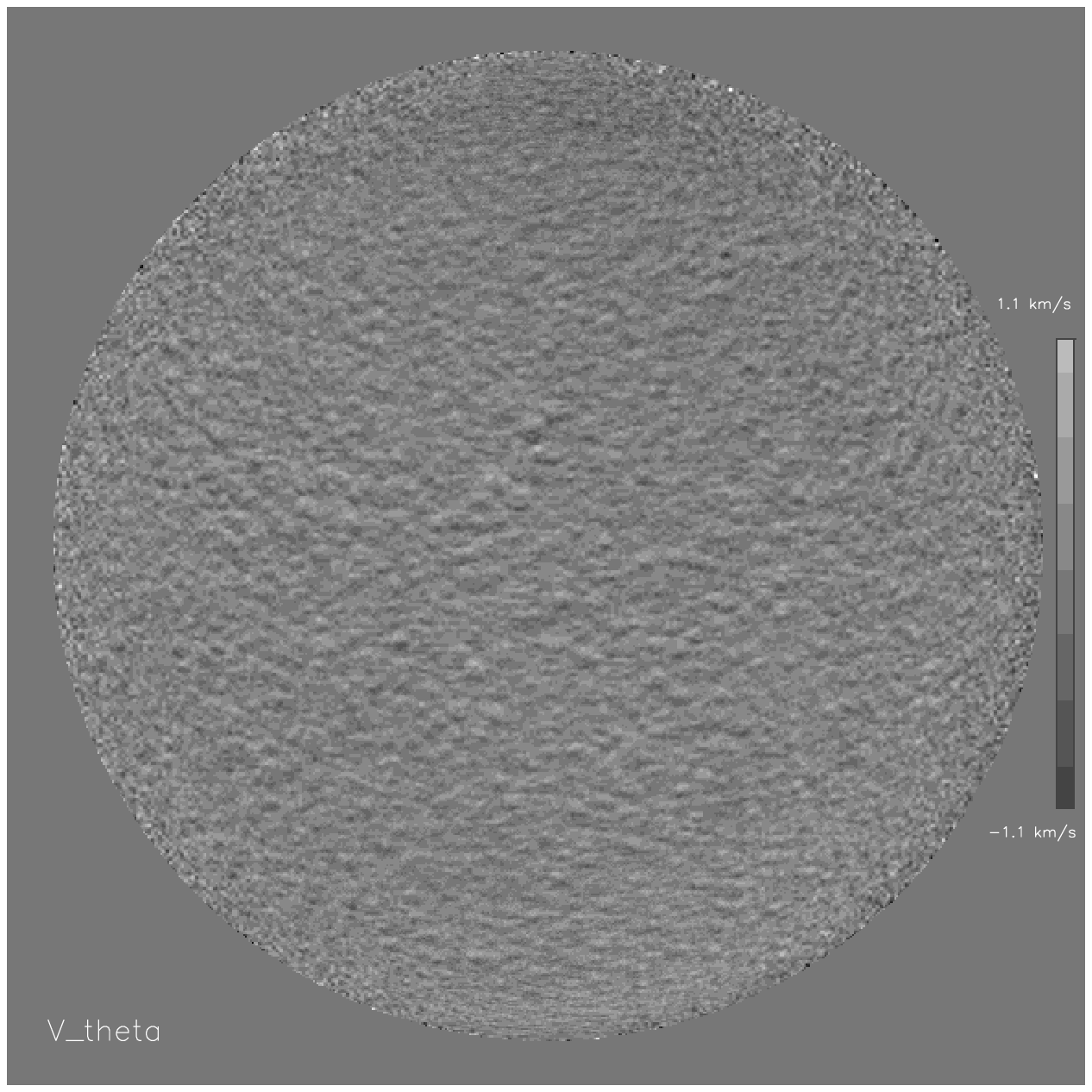}}\\
\end{center}
\caption[]{$ v_r(\theta,\varphi), v_\theta(\theta,\varphi),
v_\theta(\theta,\varphi)$ for the three-hour sequence on August 30, 2010.}
\label{spheric}
\end{figure}

Figure~\ref{spheric} shows the resulting velocities  $v_r, v_\theta,
v_\varphi$ for the three-hour sequence on August 30, 2010. The $v_r$ map
exhibits the radial component where the downflow is visible in the
sunspot regions. Close to disk center we observe a lower contrast
of $v_r$, due to the low contrast in Doppler
velocity in this part of the Sun because of the mostly horizontal flow on 
the Sun's surface. The $v_r$ component is not well determined close to the 
limb because the projection effects are predominant. The $v_\varphi$ component 
clearly shows the solar differential rotation with a lower velocity close to 
the pole. To our knowledge, this is the first time that we can measure and 
visualize the solar differential rotation with a three-hour time sequence. The
latitudinal component $v_\theta$  map allows us to essentially observe
motion on supergranulation scales with a comparatively short
time sequence.

\begin{figure*}
\resizebox{8cm}{!}{\includegraphics{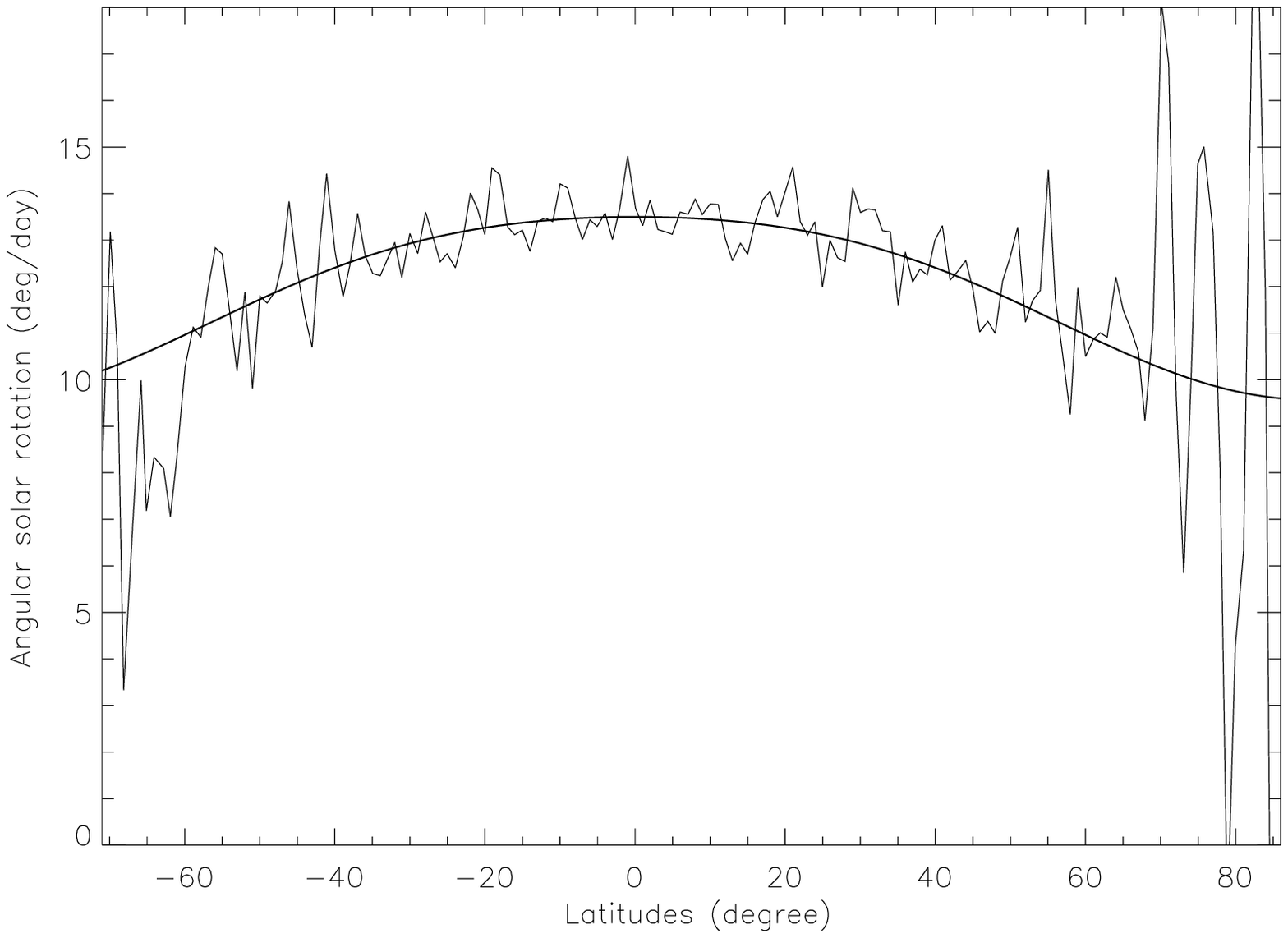}}
\resizebox{8cm}{!}{\includegraphics{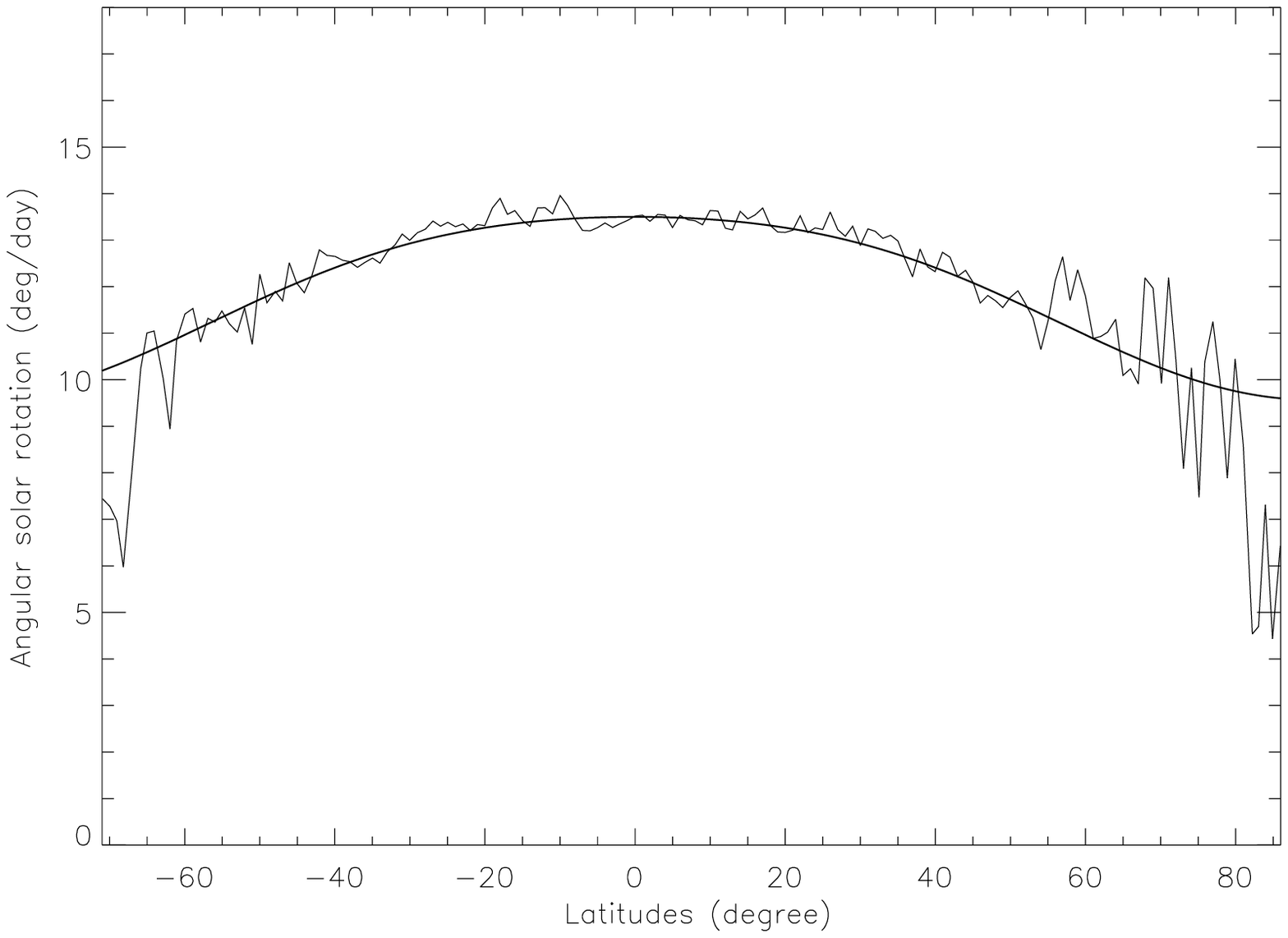}}\\
\resizebox{8cm}{!}{\includegraphics{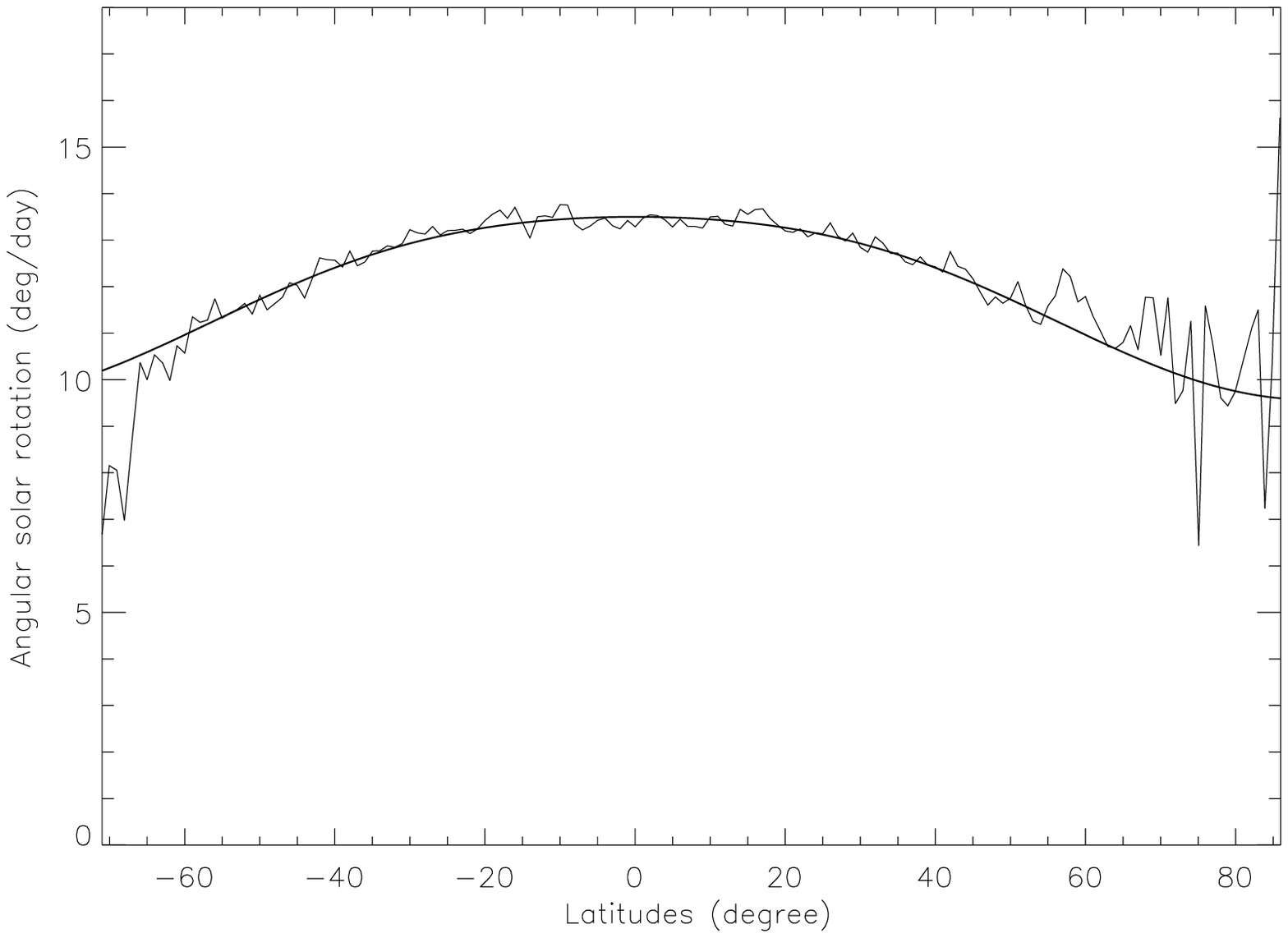}}
\resizebox{8cm}{!}{\includegraphics{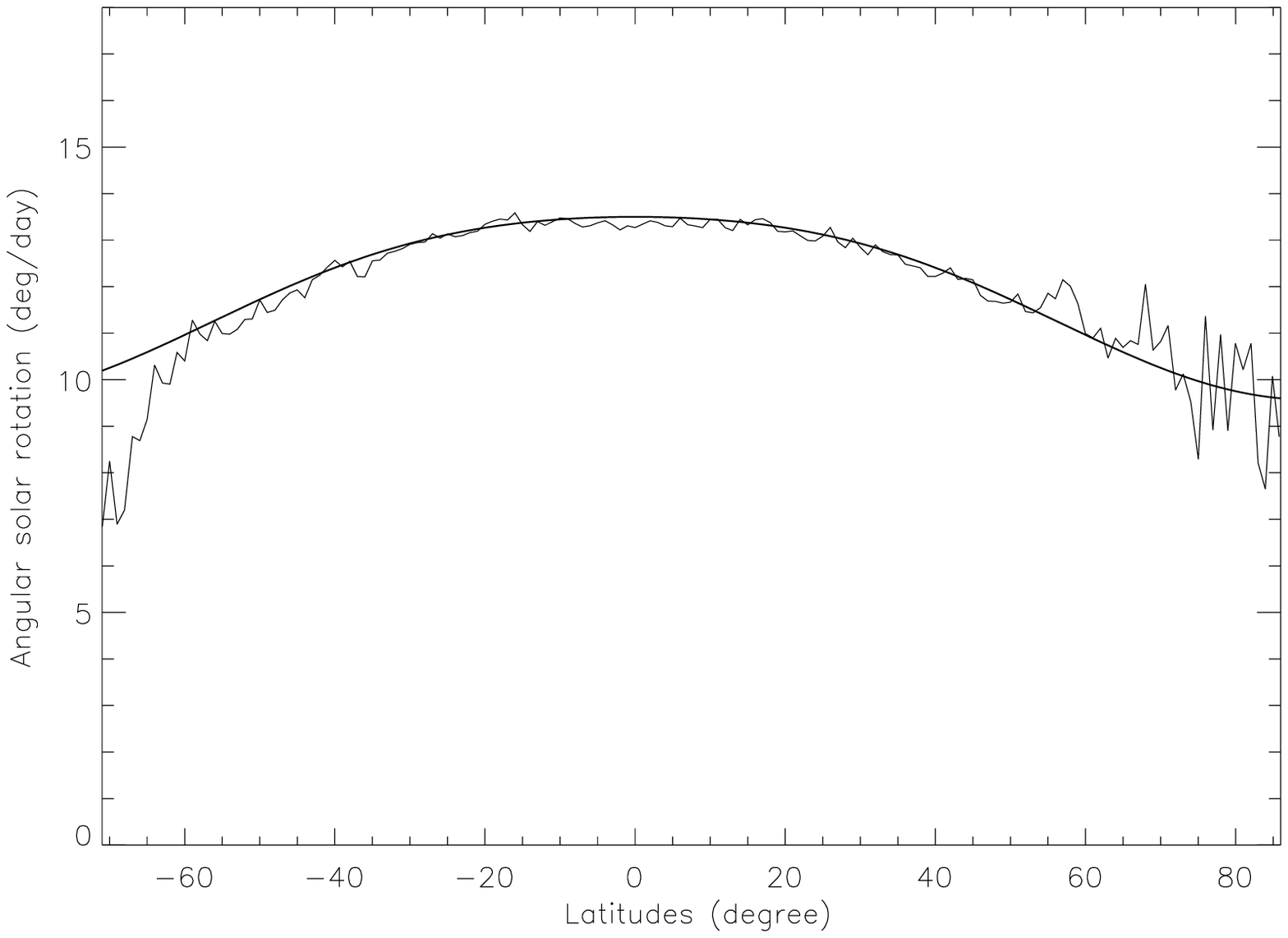}}\\
\caption[]{Solar differential rotation on  August 30, 2010, computed for
the four different angles: 2\degr (top left), 10\degr (top right),
20\degr (bottom left), 40\degr (bottom right) either side of the central
meridian.}
\label{rotdif}
\end{figure*}

\section{Determination of the solar differential rotation}

One of the first scientific applications of the CST algorithm on
SDO/HMI data described in \cite{Roud2012} was  to determine the solar
rotation  from the granule displacements. From the longitudinal velocity
component $v_\varphi$, it is possible to calculate the solar differential
rotation. In order to reduce the noise, we average $v_\varphi$ over
bands in longitudes.  Figure~\ref{rotdif} shows the computed differential
rotation averaged over bands limited to longitude of $(-2,2)$, $(-10,10)$,
$(-20,20)$, $(-40,40)$ degrees. The overplotted continuous line
represents the solar rotation measured by the spectroscopic method
\cite[][]{HH70}, which is our reference to evaluate the noise of our
measurements. The noise level is found to be 0.94, 0.37, 0.27, 0.267
\degr/day for the set of above-mentioned longitudinal bands. Thus, the
average equatorial solar rotation determined for these four bands around
the central meridian is 1.99~ km~s$^{-1}$ $\pm$ (0.133, 0.052, 0.038,
0.037)~ km~s$^{-1}$ for the different longitudinal bands.

Our results show that, for the first time, using the CST we can get a
determination of the solar rotation with a time sequence as short as three hours 
at very high precision, namely, 1.9\% in the best case.

\section {Discussion and conclusion}

The comparison between high and low spatial resolution observations
from the Hinode and SDO satellites allowed us to quantify the quality of the
horizontal velocities determined over the full Sun. We
found that the differential rotation along the central meridian is
well determined up to latitude  60\degr. With a longer time series, 
higher latitudes will be reached on depending the $B_0$ angle. Comparison 
of the meridional components shows an offset of 0.4~km~s$^{-1}$ due to
a combination of three factors: the low spatial resolution, the limb
gradient contrast, and the segmentation process. However, we describe
a way to measure and correct for the radial effect all over the Sun and
get nearly a full Sun velocity measurement. From the velocities $v_{\rm
x}$ and $v_{\rm y}$  measured in the sky plane and the simultaneous
line of sight velocity from SDO/HMI dopplergrams, we  derived the
spherical velocity components ($v_r, v_\theta, v_\varphi$). From the
longitudinal component, it is possible to get the solar differential
rotation with high precision ( $\pm 0.037$~ km~s$^{-1}$) using a temporal
sequence of only three hours. This is remarkable because other methods
require at least one month of data. That particularity opens a new field
of study of the solar rotation and motions over the solar surface. In this way, 
we can revisit the dynamics of the solar surface at high spatial
and temporal resolution from hours to months and years with the SDO data.
 In particular, it will be of great interest to compare CST convective 
velocities with numerical simulations and constraints from helioseismology 
\cite[][]{Gizon2012} .

\begin{acknowledgements}
We thank the \textit{Hinode} team for letting us use their data.
\textit{Hinode} is a Japanese mission developed and launched by
ISAS/JAXA, collaborating with NAOJ as a domestic partner and with NASA and STFC
(UK) as international partners. Support for the post-launch operation
is provided by JAXA and NAOJ (Japan), STFC (U.K.), NASA, ESA, and NSC
(Norway).  We thank the HMI team members for their hard work.We thank the
German Data Center for SDO and BAS2000 for providing SDO/HMI data.We thank
F. Rincon for his private comments.  This work was granted access to the
HPC resources of CALMIP under the allocation 2011-[P1115]. L.G. acknowledges 
support from DFG Collaborative Research Center 963 "Astrophysical Flow Instabilities 
and Turbulence" (Project A1). This work was supported by the CNRS Programme National 
Soleil Terre. M.~\v{S} is supported by the Czech Science Foundation (grant P209/12/P568).
\end{acknowledgements}

\bibliographystyle{aa}
\bibliography{biblio}

\end{document}